\documentclass[showpacs,aps,pre,preprint,superscriptaddress,textsuperscript]{revtex4-1}

\usepackage{graphicx}
\usepackage{amssymb,mathrsfs,amsmath}
\usepackage[sort&compress]{natbib}
\usepackage{cases}
\usepackage{textcomp}
\allowdisplaybreaks

\usepackage[colorlinks=true, linkcolor=blue, anchorcolor=blue, citecolor=blue, urlcolor=blue]{hyperref}

\begin{document}

\title{Density-driven segregation of binary granular mixtures \\ in a vertically vibrating drum: the role of filling fraction}

\author{Anghao Li} \author{Zaizheng Wang} \author{Haoyu Shi} \author{Min Sun}
\author{Decai Huang}
\email[Corresponding author:] {hdc@njust.edu.cn}
\affiliation{Department of Applied Physics, Nanjing University of Science and Technology, Nanjing 210094, China}

\date{\today} 

\begin{abstract}

  This paper investigates the influence of filling fraction on segregation patterns of binary granular mixtures in a vertically vibrating drum through experiments and simulations. Glass and stainless steel spherical grains, which differ in mass density, are used to form density-driven segregation. The results reveal four segregation patterns, including the Brazil-nut-effect (BNE) segregation, counterclockwise two-eye-like segregation, dumpling-like segregation, and clockwise two-eye-like segregation. The theoretical analysis demonstrates that grains predominantly exhibit counterclockwise convection at low filling fractions, while clockwise convection dominates at high filling fractions. The competition between buoyancy and convection forces determines the final stable segregation pattern. These findings provide valuable insights into controlling segregation in granular systems, which is crucial for optimizing industrial processes in fields such as pharmaceuticals and chemical engineering.
  
\end{abstract}


\date{\today}

\maketitle
\section{Introduction}
\label{Intro}

Granular materials, consisting of substantial discrete grains, are ubiquitously observed in nature and daily life \cite{Jaeger1996, Aranson2006, Mark2018FM50.407, Guo2015FM47.21, Durian2015NC6.6527}. External excitations, such as gravity \cite{Huang2024AIChE70.e18583, Huang2022PT401.117271}, vibration \cite{Hou109.198001PRL2012, Huang69.e18101AIChEs2023}, and shear \cite{Lueptow97.062906PRL2018, Shi103.052902PRE2021}, must be applied to granular systems due to the dissipative collisions between grains. Granular systems can exhibit complex segregation behaviors, where binary mixtures may spontaneously separate due to differences in density, size, and surface roughness. This segregation behavior is of great importance in process treatments such as drying and heat transfer within the pharmaceutical and chemical industries \cite{CEJ465.142756Y2023, CEJ409.128039Y2021, PRL118.218001Y2017, PT387.205Y2021}. Although significant progress has been made, the underlying physical mechanisms driving segregation remain intriguing, and controlling the segregation patterns of binary mixtures requires deeper understanding for both theoretical investigation and practical application.

External energy has to be continuously pumped into granular systems to keep grains flowing due to the intrinsic dissipation of granular interactions. Vertical vibration is a common method, leading to a widely acknowledged segregation pattern, the Brazil-nut effect (BNE) \cite{PRL90.014302Y2003, PRL86.3423Y2001, Schroter2006PRE74.011307}. In a binary granular system, smaller grains are more easily captured by gaps compared to larger grains, leading to smaller grains depositing at the bottom layer while larger grains float to the top layer. As a counterpart to BNE segregation, density-driven segregation is another typical collective motion where grains with higher mass density prefer to sink compared to lighter grains \cite{Lueptow2020PRF5.044301, Shi2007PRE75.061302, Dong2022APT33.103551}. Percolation and buoyancy effects are considered the intrinsic driving forces that determine size-based and density-driven segregation patterns, respectively. The segregation degree becomes more significant when both percolation and buoyancy act favorably on segregation. Grains with smaller size and higher mass density move easily toward the bottom, while those with larger size and lower mass density are shifted toward the top layer. On the contrary, a reverse segregation pattern can occur when the two driving forces act in opposite directions in a binary system with smaller, lower-density grains and larger, higher-density grains. The final stable segregation and mixing pattern depends on the outcome of the competition between the percolation effect and the buoyancy effect.

Based on these basic segregation and mixing driven mechanisms, various external excitations, such as self-gravity and vertical vibration, were employed to achieve targeted segregation and mixing states \cite{Hou2008PRL100.068001, Lueptow2021PRF6.054301, Lueptow2019AR10.5, Oshtorjani2021PRE103.062903, Wang2021SciAdv7.8737, Schnautz2005PRL95.028001, Du2011PRE84.041307, Cai2020PRE101.032902}. Different segregation patterns have been discovered, such as strip segregation on inclined slopes \cite{Valance2022JFM935.A41, Jenkins2015JFM782.405, Gray2012JFM709.543} and ring-like segregation in rotating drums \cite{Hsiau2020APT31.94, Huang2013EPJE36.41, Huang2012PRE85.031305}. Generally, grains with different properties in size and mass density complete the segregation or mixing process during flow. Compared to normal classical fluids, an external driving force serves as a heat source. Granular temperature, defined by the kinetic energy of grains, is introduced to characterize the dynamic behavior of granular systems. The special intrinsic nonlinear interactions between grains lead to a nonuniform distribution of granular temperature. Higher granular temperature corresponds to larger separation gaps among grains, creating local regions of reduced packing density.
When gradients in granular temperature and mass density are sufficiently large, similar to natural convection in fluids, the buoyancy force can induce instability and drive natural convection within granular systems \cite{JFM972.A29, APT35.104455Y2024, Hsiau2000CES55.3627, Yule2013PRL111.038001}. The interaction between grains differs from that between grains and container walls, creating gradients in granular temperature. Altering the properties of container walls to achieve specific convection patterns is another important method for segregating binary granular mixtures. Unlike buoyancy-driven convection, the driving force for granular flow is primarily due to the motion of the container walls. This type of convection, analogous to forced convection in fluids, is driven by the movement of the container walls rather than buoyancy. The buoyancy force drives the density-driven segregation of binary granular mixtures during the flowing process.

In vertical vibrating rectangular containers with vertical walls, two distinct convection modes of grains have been observed: upward and downward modes \cite{PRL85.1230Y2000, PRL117.098006Y2016, PRE54.874Y1996, IJMS255.108472Y2023, AIChE48.1430Y2002, PT184.166Y2008}. In the upward mode, grains flow upward in the middle of the container and downward along the vertical side walls, whereas in the downward mode, the convection direction is reversed, with grains flowing downward in the middle and upward along the vertical side walls. This transition between downward and upward convection modes depends closely on the shear force applied to the grains from the side walls. The upward convection mode tends to occur when there is less friction between the grains and the side walls. Conversely, increased friction between the grains and the side walls promotes the downward convection mode. Recently, in vertical vibrating drums, higher granular temperatures were found near the lateral walls compared to regions farther from them \cite{Huang2024AIChE70.e18583}.
The gradient in granular temperature is higher in vertical vibrating drums than in vertical vibrating rectangular containers. The reversal of density-driven segregation starts from the lateral walls and then extends to the central regions. In this study, the filling fraction is kept at a low value. Building on this work, the role of filling fraction in density-driven segregation is extensively investigated in this paper. The driving force from the curved walls on the granular system differs significantly between low and high filling fractions.

This study aims to explore how filling fraction affects the segregation patterns of binary grains with different densities in a vertically vibrating drum. Section \ref{SimMod} details the experimental setup and simulation model. In Section \ref{Results}, four segregation patterns are presented, alongside the temporal evolution of mass center and grain trajectory plots. By analyzing grain-drum wall collisions, a theoretical explanation is provided regarding the competition between buoyancy and convection forces. Quantitative flow rate results at the drum's central axis support this theory. Finally, Section \ref{Conclusions} summarizes the study's findings.

\begin{figure}[htbp]
	\centering
	\includegraphics[width=0.45\textwidth,trim=0 0 0 0,clip]
	{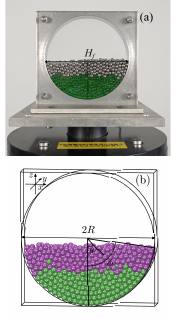}
	\caption{Snapshots of the system at the beginning. (a) Experimental setup with filling height $H_f$, and the corrsponding filling fraction $\Phi$ can be calcullated directly, (b) Simulation system with radius $R$. The coordinate origin is placed at the center of the bottom for all experiments and simulations. }
	\label{fig:FigModelExpSim}
\end{figure}

\section{Experimental setup and simulation model}
\label{SimMod}

Fig. \ref{fig:FigModelExpSim}(a) shows the experimental setup, which includes a vertical vibrating shaker and a drum. The drum's frame is constructed from aluminum alloy with a diameter \(2R = 100~{\rm mm}\) and a thickness \(T_c = 20~{\rm mm}\). The front and rear sides of the drum are covered by two plastic plates. The drum is placed on a vertical vibration shaker, which provides a sinusoidal displacement in the \(z\) direction, described by \(z_{\text{vib}} = A \sin(2\pi ft)\), where \(A = 1.5~{\rm mm}\) and \(f=20~{\rm Hz}\) are the vibration amplitude and frequency, respectively. The vibration period is \(T = 1/f\), and the maximum vibration velocity is \(v_0 = 2\pi Af\). In the experiments, stainless steel and glass spheres are used as the heavy \((h)\) and light \((l)\) grains, respectively. An initial two-layer configuration exhibiting reversal BNE (RBNE) segregation is prepared as shown in Fig. \ref{fig:FigModelExpSim}(a). In this configuration, glass grains are first randomly poured into the container, followed by stacking stainless steel grains on the top layer. \(H_f\) is the filling height within the range \([0, 2R]\). The filling angle and the filling fraction are defined by the relation \(\Phi = |\phi|/\pi\) in the range \(\phi \in [-\pi, \pi]\) and \(\Phi \in [0, 1]\), respectively.

The same simulation system is used, as shown in Fig. \ref{fig:FigModelExpSim}(b), where the glass and steel grains are initially positioned at the bottom and top layers, respectively. The vibration acceleration is defined as \(\Gamma = A(2\pi f)^2/g\), where \(g = 9.8~{\rm m/s^2}\) is the gravitational acceleration. In this work, \(f\) and \(\Gamma\) are set to \(20~{\rm Hz}\) and \(2.4\), respectively. In the simulations, the grain motion is described using Newton's equations, as detailed in our previous works \cite{Huang2006PRE74.061301, Huang2012PRE85.031305, Huang2013EPJE36.41, Huang2022PT401.117271, Huang2024AIChE70.e18583}. The acceleration due to gravity is set to \(g = 9.8~{\rm m/s^2}\), consistent with experimental conditions. The positions and velocities of the grains are updated using the Verlet algorithm at each simulation time step. A soft sphere model is adopted to describe the interaction between two contacting grains, wherein both normal and tangential forces are considered. The normal interaction at the contact point is described using the Cundall-Strack formula \cite{Cundall1979, Schafer1996, Huang2006PRE74.061301}:
\begin{equation}
	F_{n}={\frac{4}{3}}E^{*}{\sqrt{R^{*}}}{\delta_{n}}^{3/2}
	-2{\sqrt{\frac{5}{6}}}{\beta}{\sqrt{S_{n}m^{*}}}V_{n}.
	\label{CundallFn}
\end{equation}

The tangential component is taken as the minor tangential force with a memory effect and the dynamic frictional force:
\begin{equation}
	F_{\tau}=-\min({S_{\tau} {\delta}_{\tau}}-2{\sqrt{\frac{5}{6}}}{\beta}{\sqrt{S_{\tau}m^{*}}}V_{\tau},{{\mu} F_{n}}){\rm{sign}}({\delta}_{\tau}).
	\label{CundallFt}
\end{equation}

In Eqs. (\ref{CundallFn}) and (\ref{CundallFt}), $n$ and $\tau$ respectively denote the normal and tangential directions at the contact point; and $\delta_{n}$ and $\delta_{\tau}$ respectively denote the normal and tangential displacements since time $t_0$ at which contact is first established. The calculation details are as follows:
\begin{equation}
	\beta={\frac{{\rm ln}e}{{\sqrt{{\rm ln}^{2}e+{\pi}^{2}}}}},~
	{\frac{1}{m^{*}}}={\frac{1}{m_{i}}}+{\frac{1}{m_{j}}}, \\  \nonumber
\end{equation}
\begin{equation}
	S_{n}=2E^{*}{\sqrt{R^{*}{\delta_{n}}}},~
	S_{\tau}=8E^{*}{\sqrt{R^{*}{\delta_{n}}}},\\  \nonumber
\end{equation}
\begin{equation}
	{\frac{1}{E^{*}}}={\frac{1-{\nu_{i}^{2}}}{E_{i}}}+{\frac{1-{\nu_{j}^{2}}}{E_{j}}},~
	{\frac{1}{R^{*}}}={\frac{1}{R_{i}}}+{\frac{1}{R_{j}}}, \\ \nonumber
\end{equation}

\noindent where $e$ is the coefficient of restitution. The quantities $m_i$ and $m_j$ are the masses of grains $i$ and $j$ making contact, respectively, and $S_{n}$ and $S_{\tau}$ characterize the normal and tangential stiffness of the grains, respectively. $E$ and ${\nu}$ are the Young's modulus and Poisson's ratio, respectively.
In our simulations, the friction coefficient $\mu$ is fixed for both heavy and light grains, and a collision between a grain and a wall is regarded as a grain-grain collision, except that the wall has infinite mass and diameter.
Table 1 lists the values of the material parameters of the grains in the simulations.
\begin{figure}[htbp]
	\centering
	\includegraphics[width=0.75\textwidth,trim=165 520 180 125,clip]
	{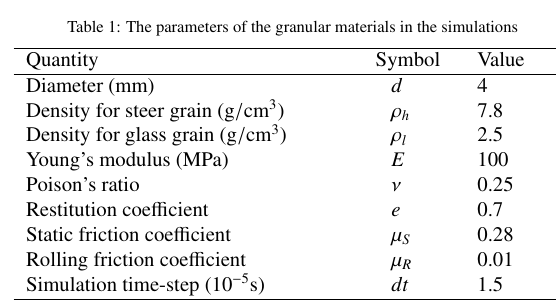}
	\label{fig:Table1}
\end{figure}

\section {Results and discussions}
\label{Results}

\begin{figure*}[htbp]
\centering
\includegraphics[width=1.0\textwidth,trim=0 0 0 0,clip]
{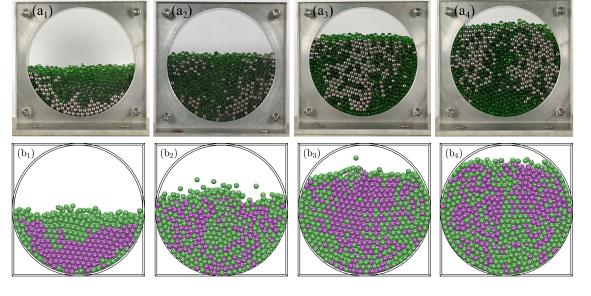}
\caption{Four typical snapshots of the steady segregation pattern for experiment and simulation results with low, mederate and high filling fractions, $({\rm a}_1)({\rm b}_1)$ $\Phi=0.48$, $N_h=600,~N_l=600$, $({\rm a}_2)({\rm b}_2)$ $\Phi=0.73$, $N_h=850,~N_l=850$, $({\rm a}_3)({\rm b}_3)$ $\Phi=0.87$, $N_h=1100,~N_l=1100$, and $({\rm a}_4)({\rm b}_4)$ $\Phi=0.95$, $N_h=1200,~N_l=1200$, respectively.
}
\label{fig:FigSegPattern}
\end{figure*}

Four typical stable segregation patterns in the experiments are shown in Fig. \ref{fig:FigSegPattern} for filling fractions \(\Phi = 0.48, 0.73, 0.87\), and \(0.95\), with the numbers of heavy and light grains being \(N_h = 600, N_l = 600\); \(N_h = 850, N_l = 850\); \(N_h = 1100, N_l = 1100\); and \(N_h = 1200, N_l = 1200\), respectively.
When the filling fraction is low (\(\Phi = 0.48\)), as shown in Fig. \ref{fig:FigSegPattern}$({\rm a}_1)$, the initial RBNE segregation pattern is reversed, leading to the occurrence of the BNE segregation pattern, where light and heavy grains migrate to the top and bottom layers, respectively. As demonstrated in Movie 1 of simulation in the Supplementary Material (SM), all grains move up and down with the vibration of the drum, and no visible convection is observed during the observation period.
Upon increasing the filling fraction to \(\Phi = 0.73\), as shown in Fig. \ref{fig:FigSegPattern}$({\rm a}_2)$, the RBNE-BNE reversal also occurs but is not stable. As shown in Movie 2 of simulation in the SM, a downward convection mode occurs. Most of the light grains flow upward along the drum walls, converging from the surface of the granular bed to the center, and then move downward along the drum center to form two symmetric closed convection rolls. The heavy grains are also involved in the convection, but most of them fluidize in the inner layers of the convection rolls.
When the filling fraction continues to increase to \(\Phi = 0.87\), as shown in Fig. \ref{fig:FigSegPattern}$({\rm a}_3)$, a dumpling-like segregation pattern is observed, where heavy grains accumulate around the center of the granular bed, while the light grains migrate toward the outer layer. As shown in Movie 3 of simulation in the SM, no visible convection is observed, and all grains only move up and down with the vibration of the container.
When the filling fraction reaches a high value of \(\Phi = 0.95\), as shown in Fig. \ref{fig:FigSegPattern}$({\rm a}_4)$, a two-eye-like segregation pattern is formed, characterized by two collective regions with heavy grains appearing symmetrically with respect to the vertical diameter. The light grains are distributed near the side wall and the central axis of the drum. As shown in Movie 4 of simulation in the SM, an upward convection mode occurs. The light grains flow downward along the drum walls, converging from the bottom of the granular bed to the center, and then move upward along the drum central line to form two symmetric closed convection rolls. The heavy grains fluidize in the inner layers of the convection rolls.

The corresponding simulations have been conducted, revealing similar segregation patterns as depicted in Fig. \ref{fig:FigSegPattern}$({\rm b}_1)$$({\rm b}_2)$$({\rm b}_3)$$({\rm b}_4)$. The correspongding movies are shown in the SM.  
No visible convection occurs for the filling fractions $\Phi=0.48$, and $0.87$. The granular bed simply moves up and down with the vibration of the drum. A stable BNE segregation pattern is observed for the lower fraction $\Phi=0.48$, whereas for the higher fraction $\Phi=0.87$, heavy grains are positioned around the center of the granular bed and light grains distribute at the outer layer.
For the filling fractions $\Phi=0.73$, and $0.95$, the downward and upward convection modes occur respectively. Two symmetric closed convection rolls appear with relative to the central axis of the drum. The light grains flow  upward (for $\Phi=0.73$)  and downward (for $\Phi=0.95$) along the drum walls, and then move downward (for $\Phi=0.73$) and upward (for $\Phi=0.95$) along the drum central line, respectively.

\begin{figure*}[htbp]
\centering
\includegraphics[width=0.9\textwidth,trim=0 0 10 0 ,clip]
{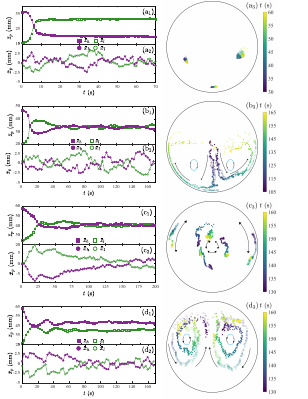}
\caption{Simulational results for temporal evolutions of the mass-center position of each grain species and typical grain trajectory for filling fractions, $({\rm a}_1)({\rm a}_2)$$({\rm a}_3)$ $\Phi=0.48$, $({\rm b}_1)({\rm b}_2)$$({\rm b}_3)$ $\Phi=0.73$, $({\rm c}_1)({\rm c}_2)$$({\rm c}_3)$ $\Phi=0.87$, and $({\rm d}_1)({\rm d}_2)$$({\rm d}_3)$ $\Phi=0.95$. The solid and open symbols represent the heavy and light grains, respectively.}
\label{fig:FigSegxTime}
\end{figure*}

Since the filling fraction has a great influence on the final stable segregation pattern, it is valuable to quantitatively characterize the dynamical process of each grain species. The mass-center of grains in the \(z\) and \(x\) directions is first introduced as in our previous work \cite{Huang2024AIChE70.e18583, Huang2013EPJE36.41, Huang2012PRE85.031305}, defined as \(\bar{z}_p = \frac{1}{N_p} \sum_{i=1}^{N_p} z_i\), where \(p = h, l\) represents the species of grains. \(N_p\) is the number of grains of species \(p\), and \(z_i\) is the height of grain \(i\) with respect to the container bottom.
Fig. \ref{fig:FigSegxTime}$({\rm a}_1)({\rm a}_2)$ shows the temporal evolution of the mass-center of grains for heavy and light grains, where the filling fraction is \(\Phi = 0.48\). At the beginning, \(\bar{z}_h\) and \(\bar{z}_l\) are large and small values, respectively, consistent with the initial RBNE pattern. When the RBNE-BNE reversal starts, \(\bar{z}_h\) decreases and \(\bar{z}_l\) increases continuously, corresponding to the sinking of heavy grains and the rising of light grains. After the segregation inversion is completed, both \(\bar{z}_h\) and \(\bar{z}_l\) stabilize at small and large values, respectively, indicating that the BNE pattern is stable. For the dynamical process of each grain species in the \(x\) direction, a change in sign indicates that heavy and light grains exchange their positions. The positive and negative signs of \(\bar{x}_h\) and \(\bar{x}_l\) indicate that they initially move toward the right and left sides of the drum, respectively. Subsequently, \(\bar{x}_h\) and \(\bar{x}_l\) change their signs and migrate toward the left and right sides of the drum. In the final steady state, both \(\bar{x}_l\) and \(\bar{x}_h\) fluctuate near zero, indicating that heavy and light grains are mixed again in the \(x\) direction.
In Fig. \ref{fig:FigSegxTime}$({\rm b}_1)({\rm b}_2)$, for \(\Phi = 0.73\), the initial RBNE is also destroyed, and heavy and light grains exchange their positions in the \(z\) direction during the initial period. When the steady state is reached, the mass centers of the heavy and light grains coincide with each other. In the \(x\) direction, the mass centers of both heavy and light grains remain near zero with some fluctuation. Similar evolutions are observed for the center positions of heavy and light grains as the filling fraction increases to \(\Phi = 0.87\) and \(0.95\). In the \(z\) direction, the initial RBNE cannot be maintained, and the mass centers of heavy and light grains reach the same stable value. The temporal evolution of the mass centers of heavy and light grains shows similar behavior, fluctuating around zero.

\begin{figure*}[htbp]
\centering
\includegraphics[width=0.85\textwidth,trim=0 0 0 0,clip]
{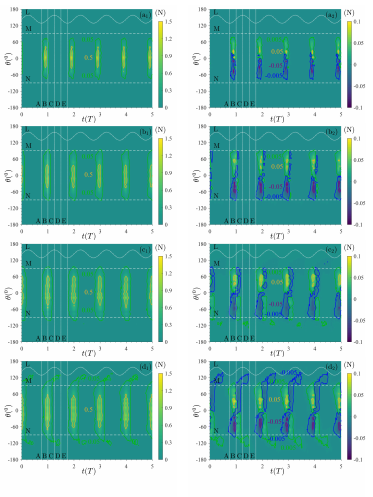}
\caption{Simulational results of the collision force between the grains and the drum wall in normal and tangential directions. The filling fractions in  $\rm (a)(b)(c)(d)$ are $\Phi=0.48$, $0.73$,$\Phi=0.87$ and $0.95$, respectively. The subscripts of $1$ and $2$ represent the driving forces  on the grains of the adjacent lay of drum wall in the normal and tangetial directions, respectively. Line The dotted $\rm L$ represents the drum vibration in phase with time. Lines $\rm M$ and $\rm N$ represent $90^0$ and $-90^0$, respectively. Lines ${\rm A,B,C,D}$ and $\rm E$ represent the times of $\frac{3}{4}T,T,\frac{5}{4}T,\frac{3}{2}T$, and $\frac{7}{4}T$, respectively. 	} 
\label{fig:FigFnFt}
\end{figure*}

The experimental and simulation results have demonstrated that variations in filling fraction lead to different segregation patterns. The collisions between grains are dissipative, and what are the active driving forces for grain motion? Obviously, the self-gravity of the grains and the vibration of the drum are two primary energy sources for the granular system. The former generates a buoyant force due to the mass density difference between heavy and light grains, leading to segregation in the direction of gravity during the flowing process. The latter can drive the granular system into specific upward and downward convection modes. Figure \ref{fig:FigFnFt} shows the simulation results of the driving forces on the grains adjacent to the drum wall in the normal and tangential directions. The positive direction of the normal direction is inward along the vertical direction of the drum wall, and the tangential direction is positive in the counterclockwise direction.
Every data point is averaged over at least 500 vibration periods at the same phase angle when a stable segregation pattern is reached. The initial times used are \(t = 30, 105, 130\), and \(130~{\rm s}\), which are the same as those in Fig. \ref{fig:FigSegxTime}. The lines M and N correspond to the filling angles of \(\phi = 90^\circ\) and \(-90^\circ\), representing the filling height \(H_f = R\) and the filling fraction \(\Phi = 0.5\).
All panels show that grains experience a periodic driving force synchronized with the drum's vibration period. For the lower filling fraction \(\Phi = 0.48\), the normal collision force is primarily confined between lines A (\(t = \frac{3}{4}T\)) and B (\(t = T\)), as shown in Fig. \ref{fig:FigFnFt}$({\rm a}_1)$. During other phases, the minimal collision force indicates that most grains are detached from the drum wall. The tangential collision force, which is the dynamic friction force \(f_\mu\), is determined by the normal collision force and the tendency of motion, as evidenced by Fig. \ref{fig:FigFnFt}$({\rm a}_2)$. It is strictly confined between lines A and B at the beginning of the ascending phase. At the right (\(\theta > 0\)) and left (\(\theta < 0\)) sides of the drum, the friction forces are positive and negative, respectively, between lines A and B. This indicates that grains experience counterclockwise and clockwise driving forces. The tangential collision force is minimal during other phases, suggesting that gravity dominates the motion of the grains. In Figs. \ref{fig:FigFnFt}$({\rm b}_1), ({\rm b}_2), ({\rm c}_1)$, and $({\rm c}_2)$, increasing the filling fraction extends the range of significant normal collisions to the phase between lines B and C. The range of the tangential collision force also extends with increasing filling fraction and changes sign between lines B and C, though it remains confined between lines M and N.
When the filling fraction is very high (\(\Phi = 0.95\)), as shown in Fig. \ref{fig:FigFnFt}$({\rm d}_1)({\rm d}_2)$, both the normal and tangential collision forces extend into the descending phase between lines C and D and into the regions \(\theta > 90^\circ\) and \(\theta < -90^\circ\). This indicates that collisions with the upper drum wall play a significant role in the forced motion of the grains.

\begin{figure*}[htbp]
\centering
\includegraphics[width=0.99\textwidth,trim=8 0 8 0,clip]
{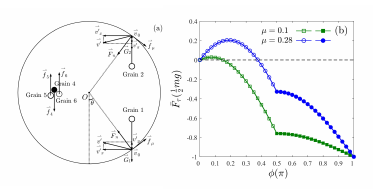}
\caption{(a) Schematic diagram of convection and buoyant forces. ${\mathop{v}\limits^{\rightharpoonup}}_0$ and ${\mathop{v'}\limits^{\rightharpoonup}}_0$ are the incident and reflection velocities, respectivley. ${\mathop{v'}\limits^{\rightharpoonup}}_x$ and ${\mathop{v'}\limits^{\rightharpoonup}}_y$ are the reflection velocities in $x$ and $y$ directions, respectively. ${\mathop{G}\limits^{\rightharpoonup}}$  is the self gravity. ${\mathop{F}\limits^{\rightharpoonup}}_n$  is the contact force in normal direction. ${\mathop{f}\limits^{\rightharpoonup}}_{\mu}$ is the friction force between the grain and the drum wall. ${\mathop{f}\limits^{\rightharpoonup}}_i$ $(i=4,5~{\rm and}~6)$ is the buoyant force for heavy and light grains. (b) Dependence of the driving force \(\bar{F}_{\tau}\) on the filling angle. Squares and  circles represent friction coefficients $\mu=0.1$ and $0.28$, respectively. Open and solid symbols represent the filling angles $\phi<\pi/2$ and $\phi>\pi/2$, respectively.
} 
\label{fig:FigCollModel}
\end{figure*}

The analysis of collisions between grains and the drum wall shows that larger collision forces in both the normal and tangential directions dominate the collision process when the filling fraction \(\Phi < 0.5\), leading to synchronized periodic motion with the drum. However, a higher filling fraction promotes collisions between grains and the upper drum wall in the region \(\Phi > 0.5\). The curved wall of the drum has a significantly different impact on collision forces when the filling fraction is increased from a lower to a higher value. Let us first consider two special collisions for grains adjacent to the drum wall at positions \(H_f < 0.5\) and \(H_f > 0.5\), as shown in Fig. \ref{fig:FigCollModel}(a). For grain 1, located at the lower region of the drum, the direction of the self-gravity ${\mathop{G}\limits^{\rightharpoonup}}$ is opposite to that of the collision force ${\mathop{F}\limits^{\rightharpoonup}}_n$ and the friction force ${\mathop{f}\limits^{\rightharpoonup}}_{\mu}$. However, for grain 2, located at the upper region of the drum, ${\mathop{G}\limits^{\rightharpoonup}}$, ${\mathop{F}\limits^{\rightharpoonup}}_n$, and ${\mathop{f}\limits^{\rightharpoonup}}_{\mu}$ contribute to downward motion.
Under the combined effect of self-gravity and tangential collision forces, grains adjacent to the drum wall exhibit two distinct forced motions: clockwise and counterclockwise, depending on the filling fraction. This directed motion is reasonably termed forced convection, analogous to fluid dynamics. When a grain collides with the drum wall, it also experiences collisions with other grains in the system. The resulting acceleration can be considered equivalent to the drum's acceleration. Additionally, the granular system experiences an upward lifting force from the drum wall during the upward motion phase, and the driving force reverses direction to become a downward pushing force during the downward motion phase. The upward motion period \(T_{\rm U}\) is equal to the downward motion period \(T_{\rm D} = T_{\rm U} = T/2\).
Therefore, considering the grain motion located at the right side of the drum, the driving force for grains adjacent to the drum wall in the tangential direction contributes to the forced motion over a vibration period and can be expressed as follows:
\begin{subequations}
\begin{align}
	\bar{F}_{\tau} 
	&=\frac{1}{T}[\int_{0}^{T_{\rm U}}\int_{0}^{\phi}{\mu}mg{\Gamma}{\cos}{\theta} d{\theta}dt 
	-\int_{0}^{T_{D}}\int_{0}^{\phi}mg{\rm sin}{\theta}d{\theta}dt], ~ & 0 \leq  {\phi} \leq \frac{1}{2}\pi, \nonumber \\
	&= \frac{1}{2}mg({\mu}{\Gamma}{\rm sin}{\phi}+{\rm cos}\phi-1). ~ &  \label{eq:Fca} \\
	\bar{F}_{\tau} 
	&=\frac{1}{T}[\int_{0}^{T_{\rm U}}\int_{0}^{\pi/2}{\mu}mg{\Gamma}{\cos}{\theta} d{\theta}dt 
	-\int_{0}^{T_{D}}\int_{0}^{\pi/2}mg{\rm sin}{\theta}d{\theta}dt ~ &		\nonumber  \\
	&-\int_{0}^{T_{\rm D}}\int_{\pi/2}^{\phi}{\mu}mg{\Gamma}{\cos}{\theta} d{\theta}dt], ~ & \frac{1}{2}\pi < \phi \leq \pi ,  \nonumber \\
	&=\frac{1}{2}mg\{{\mu}{\Gamma}-[1+{\mu}{\Gamma}(1-{\rm sin}{\phi})]\} &  \nonumber \\
	&=\frac{1}{2}mg({\mu}{\Gamma}{\rm sin}\phi-1). ~ & \label{eq:Fcb}
\end{align}
\label{eq:Fc00}
\end{subequations}

In Fig. \ref{fig:FigCollModel}(b), the dependence of the driving force \(\bar{F}_{\tau}\) on the filling angle is plotted for two friction coefficients between the grains and the drum wall, \(\mu = 0.1\) and \(0.28\). The grains are forced to move counterclockwise for positive \(\bar{F}_{\tau}\), while the grains flow clockwise for negative \(\bar{F}_{\tau}\). Obviously, increasing the friction coefficient between the grains and the drum wall favors counterclockwise motion.
Increasing the filling fraction is equivalent to increasing the contact angle between the grains and the drum wall. Negative \(\bar{F}_{\tau}\) indicates the formation of clockwise motion of the grains. These characteristics are consistent with the experimental results \cite{AIChE48.1430Y2002}.
At the same time, the grains experience the buoyant effect, as demonstrated by grains 4, 5, and 6. The heavy grain 4 sinks downward, while light grains 5 and 6 migrate upward. The corresponding buoyant force for a single grain can be calculated similarly to that in fluids.

\begin{equation}
F_{\rm B} = {\rho_{\rm eff}}g{V_0}
\label{eq:F_B}
\end{equation}

\noindent where $\rho_{\rm eff}={(\rho_{h}+\rho_{l})/2}$ is the effective mass density of the granular system. 

Based on the features of convection force and buoyancy force, the segregation phenomena observed due to variations in filling fraction can be attributed to their competitive effects. The former promotes forced convection behavior in the granular system, while the latter favors the occurrence of BNE segregation. Here, only the right half of the drum is considered due to symmetry. At lower filling fractions, buoyancy has a much stronger influence on the segregation process than convection. The granular system eventually reaches a stable BNE segregation pattern, where heavy grains accumulate at the bottom layer and light grains gather at the top layer, as shown in Fig. \ref{fig:FigSegPattern}$(\rm a_1)(\rm b_1)$. An increase in filling fraction below one-half enhances the counterclockwise convection force, which eventually surpasses the buoyancy force. The granular system enters a fully counterclockwise convection flow. Grains flow upward along the drum wall and then downward along the central axis of the drum. However, heavy grains sink due to the buoyancy effect and accumulate at the center of the convection cell, as shown in Fig. \ref{fig:FigSegPattern}$(\rm a_2)(\rm b_2)$. When the filling fraction exceeds one-half, clockwise convection begins to dominate. For a filling fraction of \(\Phi = 0.87\), the effects of clockwise and counterclockwise convection, as well as buoyancy, are all significant. The resulting convection flow halts, causing heavy grains to gather at the drum's center and light grains to remain on the periphery, as shown in Fig. \ref{fig:FigSegPattern}$(\rm a_3)(\rm b_3)$. When the filling fraction reaches a high value of \(\Phi = 0.95\), clockwise convection dominates, forcing all grains to flow clockwise. Simultaneously, the buoyancy effect causes heavy grains to sink first, allowing light grains to flow outward while heavy grains move inward, as shown in Fig. \ref{fig:FigSegPattern}$(\rm a_4)(\rm b_4)$. Additionally, decreasing the friction coefficient tends to increase the negative value of  
\(\bar{F}_{\tau}\), indicating that a smaller friction coefficient favors the formation of a counterclockwise convection mode. 

\begin{figure*}[htbp]
\centering
\includegraphics[width=0.65\textwidth,trim=100 270 100 250,clip]
{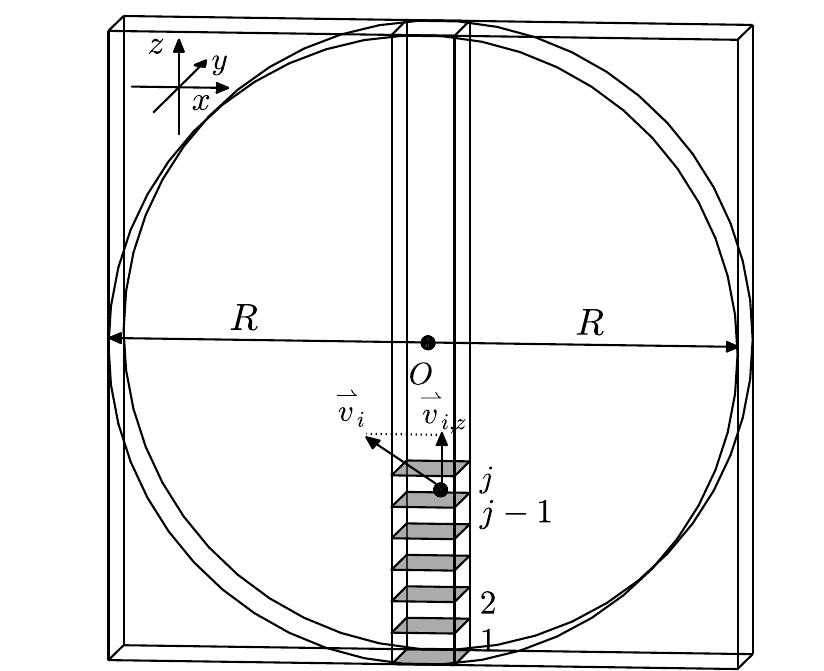}
\caption{ A schematic diagram of subregion and the contributed flow rate $q_{i}$ in the $z$ direction for subregion $j$. 
} 
\label{fig:FigBin}
\end{figure*}

\begin{figure*}[htbp]
\centering
\includegraphics[width=0.99\textwidth,trim=0 0 0 0,clip]
{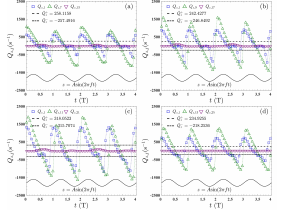}
\caption{Simulational results of local flow rate in the $z$ direction for filling fractions, $({\rm a})$ $\Phi=0.48$, $({\rm b}_1)$ $\Phi=0.73$, $({\rm c})$ $\Phi=0.87$, and $({\rm d})$ $\Phi=0.95$. The dash and dash-dot lines represent the summations of positive $\bar{Q}_{z}^{+}$ and negative $\bar{Q}_{z}^{-}$ local flow rates of all subregions, respectively. The solid line represents the vibration position of the drum.}
\label{fig:FigQTime}
\end{figure*}

To quantify the influence of filling fraction on the convection effect, subregions are selected along the central axis of the drum, as shown in Fig. \ref{fig:FigBin}. The size of each subregion is \(L_{x} = 20~\text{mm}\), \(L_{y} = 20~\text{mm}\), and \(L_{z} = 5~\text{mm}\), and a total of 25 subregions are obtained. The local flow rate in the \(z\) direction through the cross-sectional area of subregion \(j\) is easily defined as:

\begin{equation}
Q_{z,j} = \sum_{i} q_{i,z}/L_{z},
\label{eq:F_B}
\end{equation}  

\noindent where $q_{i,z}$ is the flow rate of the $i$th grain in the $j$ subregion.

In Fig. \ref{fig:FigQTime}, the time evolution of the local flow rates in three typical subregions, the bottom, middle, and top, is plotted for filling fractions $\Phi=0.48, 0.73, 0.87$ and $0.95$. For the top subregion, the local flow rate is naturally close to zero due to the sparse distribution of grains. The bottom and middle subregions exhibit periodic upward and downward flow rates synchronized with the drum's vibration.
The flow rate through the central region is obtained by summing the flow rates of all subregions and is expressed as $\bar{Q}=\bar{Q}^{+}+\bar{Q}^{-}$, where $\bar{Q}^{+}$ and $\bar{Q}^{-}$ represent the summations of positive and negative local flow rates across all subregions, respectively. A positive $\bar{Q}$ indicates upward flow at the drum's center, corresponding to clockwise convection flow, while a negative $\bar{Q}$ indicates counterclockwise convection flow. No convection flow is observed when $\bar{Q}=0$. 

As shown in Fig. \ref{fig:FigQTime}(a)(b)(c)(d), the flow rates $\bar{Q}_z$ are $0.6242,-4.4215,2.3449$ and $16.6718~{\rm s^{-1}}$ for the respective filling fractions. For filling fractions $\Phi=0.45$ and $0.87$, the flow rates are small, indicating weak convection. In contrast, higher flow rates are observed for $\Phi=0.73$ and $0.95$. The negative and positive signs of $\bar{Q}$ reveal that convection is counterclockwise for $\Phi=0.73$ and clockwise for $\Phi=0.95$. These observations align with the convection modes depicted in Fig. \ref{fig:FigSegPattern}.

\begin{figure*}[htbp]
\centering
\includegraphics[width=0.95\textwidth,trim=0 250 0 250,clip]
{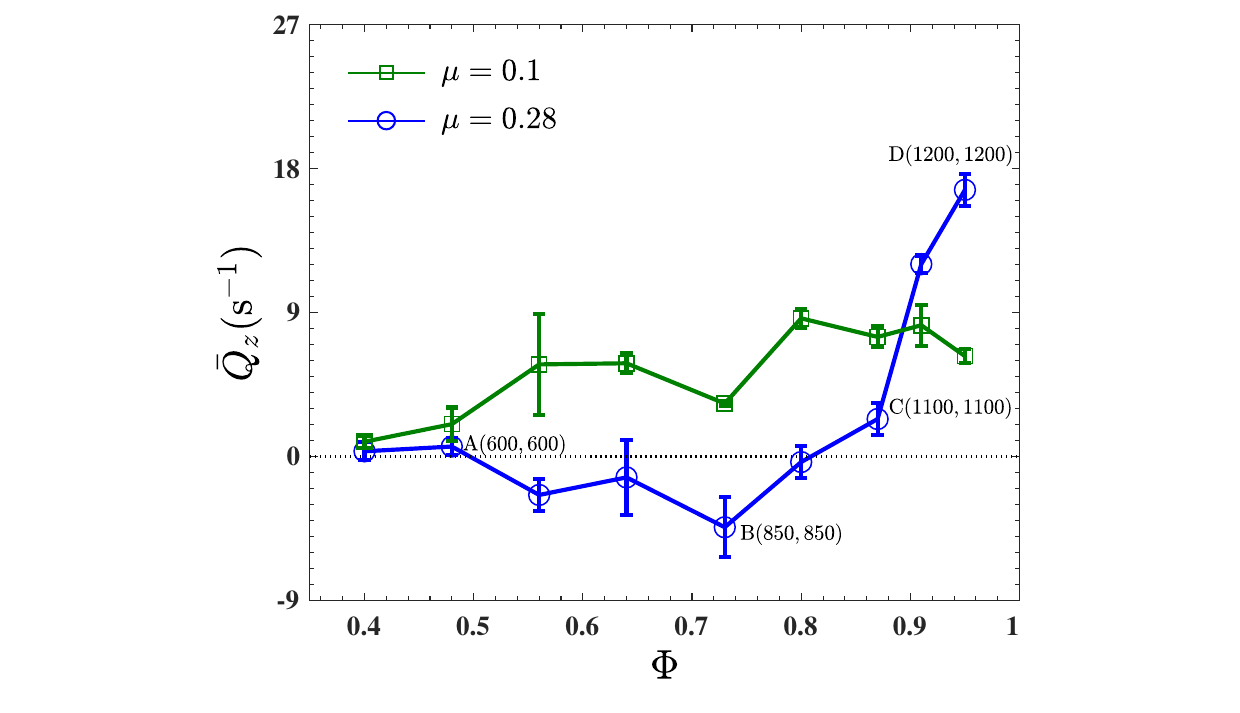}
\caption{Flow rate  $\bar{Q}_z$ as a function of filling fraction $\Phi$ for friction coefficient $\mu=0.1$ (sqarues) and $0.28$ (circles). Points A,B,C, and D corespond to filling fraction $\Phi=0.48,0.72,0.87$ and $0.95$, respectively.
} 
\label{fig:FigQPhi}
\end{figure*}

In Fig. \ref{fig:FigQPhi}, we summarize the influence of filling fraction on the convection mode in the vertically vibrating drum. The friction coefficient $\mu$=0.28 is firstly used. At low filling fractions, convection flow is minimal, and the central flow rate approaches zero. Buoyancy drives heavy grains to settle at the bottom, forming a BNE segregation pattern as shown in Fig. \ref{fig:FigSegPattern}$(\rm a_1)(b_1)$.
As the filling fraction increases, the counterclockwise convection force grows, creating a downward convection mode. Both heavy and light grains are influenced by convection, with heavy grains sinking to the center of the convection roll and light grains moving to the periphery, forming a counterclockwise two-eye-like segregation pattern shown in Fig. \ref{fig:FigSegPattern}$(\rm a_2)(b_2)$. When the filling fraction continues to rise and the convection and buoyancy effects balance, convection ceases, leading to a dumpling-like pattern where heavy grains are centralized and light grains form an outer layer, as shown in Fig. \ref{fig:FigSegPattern}$(\rm a_3)(c_3)$.
At very high filling fractions, the clockwise convection force dominates, restarting grain flow. BNE occurs at the top surface, with heavy grains sinking first, and the clockwise two-eye-like segregation pattern emerges in Fig. \ref{fig:FigSegPattern}$(\rm a_4)(d_4)$, where heavy and light grains are distributed in the inner and outer layers of the convection roll. A low friction coefficient $\mu$=0.1 is also used in the simulations. The flow rate is expected to be positive, indicating the clockwise convection mode. These characteristics are consistent with the discussion of Eq. \ref{eq:Fc00}. The clockwise convection force will increase when the friction coefficient decreases.

\section{Conclusions}
\label{Conclusions}

In this study, experiments and simulations are conducted to investigate the influence of filling fraction on segregation patterns in binary granular mixtures in a vertically vibrating drum. Glass and stainless steel particles, differing in mass density, are employed to analyze density-driven segregation dynamics. Vertical and horizontal mass centers are defined to quantify segregation intensity. Four distinct segregation patterns emerge under varying filling fractions: Brazil Nut Effect (BNE), counterclockwise two-eye-like segregation, dumpling-like segregation, and clockwise two-eye-like segregation pattern, observed at low, medium, high, and very high filling fractions, respectively.

Simulation results reveal that at low filling fractions, normal and tangential collision forces predominantly occur during the ascending vibration phase. Grain-wall collisions are localized near the drum's bottom region, where $-\pi/2<\theta<\pi/2$. As filling fraction increases, collisions extend beyond $\theta>\pi/2$ and $\theta<-\pi/2$. This collision profile indicates that at lower filling fractions, grains primarily experience upward collisions during drum ascent and gravitational settling during descent. Downward collisions critically influence segregation dynamics. Theoretical analysis demonstrates that counterclockwise convection dominates at lower filling fractions due to upward collision prevalence, while increased downward collisions at higher filling fractions induce clockwise convection. With small friction coefficients, weak convection forces allow buoyancy to dominate, where the stronger downward driving force stabilizes clockwise convection. 
The competition between convection forces and buoyancy governs filling fraction effects on segregation patterns. At low filling fractions, buoyancy-driven BNE prevails. Increasing filling fraction enhances counterclockwise convection, causing heavy grains to sink from the top flowing layer and accumulate at convection roll centers, forming a two-eye-like pattern. At intermediate filling fractions, balanced convection and buoyancy produce dumpling-like segregation (heavy grains centrally, light grains peripherally). At very high filling fractions, dominant clockwise convection overpowers buoyancy, regenerating the two-eye-like pattern. Additional simulations corroborate these mechanisms through flow rate analysis in the drum's central region. These findings hold theoretical significance and industrial relevance for optimizing granular mixture processes.

\begin{acknowledgments}
This work are financially supported by the National Natural Science Foundation of China (Grant No. 11574153) and the fund of No.TSXK2022D007.
\end{acknowledgments}

\end{document}